\documentclass[aps,showpacs,superscriptaddress,preprint]{revtex4}%
\usepackage{amsfonts}
\usepackage{amsmath}
\usepackage{amssymb}
\usepackage{graphicx}%
\begin{document}
\title{Quantum spin Hall effect and spin-charge separation in a kagom\'{e} lattice}
\author{Zhigang Wang }
\affiliation{LCP, Institute of Applied Physics and Computational Mathematics, P.O. Box
8009, Beijing 100088, People's Republic of China}
\author{Ping Zhang}
\thanks{Corresponding author. Email address: zhang\_ping@iapcm.ac.cn}
\affiliation{LCP, Institute of Applied Physics and Computational
Mathematics, P.O. Box 8009, Beijing 100088, People's Republic of
China} \affiliation{Center for Applied Physics and Technology,
Peking University, Beijing 100871, People's Republic of China}
\pacs{73.43.-f, 71.10.Pm, 72.25.Hg}
\date{\today}

\begin{abstract}
A two-dimensional kagom\'{e} lattice is theoretically investigated
within a simple tight-binding model, which includes the nearest
neighbor hopping term and the intrinsic spin-orbit interaction
between the next nearest neighbors. By using the topological winding
properties of the spin-edge states on the complex-energy Riemann
surface, the spin Hall conductance is obtained to be quantized as
$-e/2\pi$ ($e/2\pi$) in insulating phases. This result keeps
consistent with the numerical linear-response calculation and the
\textbf{Z}$_{2}$ topological invariance analysis. When the sample
boundaries are connected in twist, by which two defects with $\pi$
flux are introduced, we obtain the spin-charge separated solitons at
$1/3$ (or $2/3$) filling.

\end{abstract}
\maketitle

Over the last two decades the topological band insulators (TBIs)
have been a subject of great interest in condensed matter field
\cite{Thouless,Wenbook}. Different from the normal band insulators,
the TBIs have a prominent feature, which is the necessary presence
of gapless edge states on the sample boundaries
\cite{Halperin,Hatsugai}. An early TBI model was proposed by Haldane
\cite{Haldane}. Therein it was shown that the gapless edge states
result in a remarkable character of TBIs by showing quantum Hall
effect in the absence of an external magnetic field. Besides the
Haldane model, several other lattice models have also been proposed
to be quantum Hall TBIs, which include the two-dimensional (2D)
\cite{Ohgushi,Tail,Wang2} and three-dimensional (3D) \cite{Tak}
spin-chiral kagom\'{e} lattices, and the 3D distorted fcc lattice
\cite{Shindou, Wang3}. All the quantum Hall TBIs rely on the
breaking of time-reversal symmetry (TRS).

Recently, Kane and Mele generalized the spinless Haldane model to a
spin one by adding an intrinsic spin-orbit interaction (SOI)
\cite{Kane1,Kane2}. The TRS is conserved in the Kane-Mele model and
the gapless spin edge states in this model result in quantum spin
Hall effect. These TBIs like the Kane-Mele model keep TRS and are
different from the quantum Hall TBIs. Thus call them the quantum
spin Hall TBIs. At present the quantum spin Hall TBIs are receiving
considerable attention. One impressive example is that only one year
later after its theoretical prediction to be a quantum spin Hall TBI
in 2006 \cite{Bernevig}, it was proved in experiment that HgTe is an
actual one \cite{Koenig}.

One special character of the quantum spin Hall TBIs was recently
attributed to their spin-charge separated excitations \cite{Ran, Qi}
in the presence of a $\pi$ flux. This attribution is motivated by
the recent advance in studying 2D fractionalized quasiparticles
\cite{Lee,Hou}, and is a straightforward result when, like what Kane
and Mele \cite{Kane1} have dealt with Haldane's TBI, considering
spins of the 2D edge soliton. The separate spinon, holon and
chargeon obey Bose statistics, and the experimental measurement of
these soliton excitations would provide an undoubted verification of
the \textbf{Z}$_{2}$ topological properties of the quantum spin Hall
TBIs. At present, besides the necessity for further studies to gain
more insights into the nature of spin-charge separation and its
connection to the other topological phenomena, obviously, it is also
important to identify and study various model systems that exhibit
the phenomenon of spin-charge separation. Motivated by this
observation, as well as by the recent attention on the layered metal
oxides as possible candidates for the quantum spin Hall TBI
\cite{Shitade}, in this paper, we study the quantum spin Hall effect
and construct spin-charge separated edge solitons in a 2D kagom\'{e}
lattice. Different from the previously studied kagom\'{e} TBI
\cite{Ohgushi,Tail,Wang2,Tak,Fuj,Zhang} whearein the presence of
ferromagnetic spin chirality breaks TRS, here TRS persists and the
quantum spin Hall effect occurs due to the intrinsic SOI.
Experimentally, the physical candidates for realizing our studied
system might be $5d$ transition metal oxides with layered pyrochlore
structure \cite{Man,Singh,Mat}. The argument is that in $5d$
transition metal oxides, both the SOI and the electron correlation
become important with the same order of magnitude. As a consequence,
at high temparture the correlation-induced magnetic order can be
overcome by SOI and the nontrivial topological insulator phase is
expected to occur \cite{Shitade,Rag,Chen,Kim,Pesin}. The other
alternative way to experimentally realize our studied system is by
modulating the 2D electron with a periodic potential with kagom\'{e}
symmetry, as recently demonstrated for artificial graphene
\cite{Gib}. By using the bulk linear-response theory, as well as the
topological winding numbers of the spin-edge states on the
complex-energy Riemann surface, we obtain the spin Hall conductance
(SHC) $\sigma_{xy}^{s}$. The quantized value of $\sigma_{xy}^{s}$ is
$-e/2\pi$ ($e/2\pi$) at 1/3 (2/3) filling. Then, we construct
spin-charge separated edge solitons by introducing $\pi$ fluxes with
a method similar to that in Ref. \cite{Lee}. The quantum statistics
of these solitons is also discussed. \begin{figure}[ptb]
\begin{center}
\includegraphics[width=0.3\linewidth]{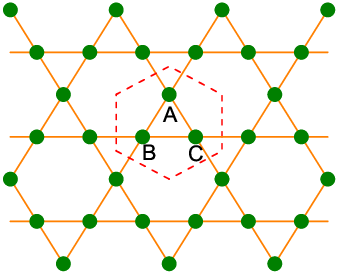}
\end{center}
\caption{(Color online). Schematic picture of the 2D kagom\'{e} lattice. The
dashed lines represent the Wigner-Seitz unit cell, which contains three
independent sites (A, B, C).}%
\label{f1}%
\end{figure}

Consider the tight-binding model for independent electrons on a 2D
kagom\'{e} lattice (Fig. \ref{f1}). The spin-independent part of the
Hamiltonian is given by
\begin{equation}
\mathcal{H}_{0}=t\sum_{\langle ij\rangle\sigma}c_{i\sigma}^{\dag}c_{j\sigma},
\label{1}%
\end{equation}
where $t_{ij}$=$t$ is the hopping amplitude between the nearest-neighbor link
$\langle ij\rangle$ and $c_{i\sigma}^{\dag}$ ($c_{i\sigma}$) is the creation
(annihilation) operator of an electron with spin $\sigma$ (up or down) on
lattice site $i$. For simplicity, we choose $t$=$-1$ as the energy unit and
the distance $a$ between the nearest sites as the length unit throughout this paper.

The Hamiltonian (\ref{1}) can be diagonalized in the momentum space as%
\begin{equation}
\mathcal{H}_{0}=\sum_{\mathbf{k}}\psi_{\mathbf{k}\sigma}^{+}H_{0}%
(\mathbf{k})\psi_{\mathbf{k}\sigma}, \label{Hk}%
\end{equation}
where the electron field operator $\psi_{\mathbf{k}\sigma}$=$(c_{A\mathbf{k}%
\sigma},c_{B\mathbf{k}\sigma},c_{C\mathbf{k}\sigma})^{\text{T}}$ includes the
three lattice sites ($A,B,C$) in the Wigner-Seitz unit cell shown in Fig. 1.
$H_{0}(\mathbf{k})$ is a $3\mathtt{\times}3$ spinless matrix given by
\begin{equation}
H_{0}(\mathbf{k})=-t\left(
\begin{array}
[c]{ccc}%
0 & 2\cos\left(  \mathbf{k}\mathtt{\cdot}\mathbf{a}_{1}\right)  & 2\cos\left(
\mathbf{k}\mathtt{\cdot}\mathbf{a}_{3}\right) \\
2\cos\left(  \mathbf{k}\mathtt{\cdot}\mathbf{a}_{1}\right)  & 0 & 2\cos\left(
\mathbf{k}\mathtt{\cdot}\mathbf{a}_{2}\right) \\
2\cos\left(  \mathbf{k}\mathtt{\cdot}\mathbf{a}_{3}\right)  & 2\cos\left(
\mathbf{k}\mathtt{\cdot}\mathbf{a}_{2}\right)  & 0
\end{array}
\right)  , \label{Hamilton}%
\end{equation}
where $\mathbf{a}_{1}$=$(-1/2,-\sqrt{3}/2)$, $\mathbf{a}_{2}$=$(1,0)$, and
$\mathbf{a}_{3}$=$(-1/2,\sqrt{3}/2)$ represent the displacements in a unit
cell from A to B site, from B to C site, and from C to A site, respectively.
In this notation, the first Brillouin zone (BZ) is a hexagon with the corners
of $\mathbf{K}$=$\pm\left(  2\pi/3\right)  \mathbf{a}_{1}$, $\pm\left(
2\pi/3\right)  \mathbf{a}_{2}$, $\pm\left(  2\pi/3\right)  \mathbf{a}_{3}$.

The energy spectrum for spinless Hamiltonian $H_{0}(\mathbf{k})$ is
characterized by one dispersionless flat band ($\epsilon_{1\mathbf{k}}^{(0)}%
$=$2$), which reflects the fact that the 2D kagom\'{e} lattice is a line graph
of the honeycomb structure \cite{Mie}, and two dispersive bands,
$\epsilon_{2(3)\mathbf{k}}^{(0)}=-1\mathtt{\mp}\sqrt{4b_{\mathbf{k}}-3}$ with
$b_{\mathbf{k}}$=$\sum_{i=1}^{3}\cos^{2}\left(  \mathbf{k\cdot a}_{i}\right)
$. These two dispersive bands touch at the corners (\textbf{K}-points) of the
BZ and exhibit Dirac-type energy spectra, $\epsilon_{2(3)\mathbf{k}}%
^{(0)}\mathtt{=}(-1\mathtt{\mp}\sqrt{3}|\mathbf{k}\mathtt{-}\mathbf{K}|)$,
which implies a \textquotedblleft particle-hole\textquotedblright\ symmetry
with respect to the Fermi energy $\epsilon_{F}\mathtt{=}-1$. The corresponding
eigenstates of $H_{0}(\mathbf{k})$ are given by%
\begin{equation}
\left\vert u_{n\mathbf{k}}^{(0)}\right\rangle =G_{n\mathbf{k}}\left(
q_{1\mathbf{k}},q_{2\mathbf{k}},q_{3\mathbf{k}}\right)  ^{\text{T}},
\label{wave}%
\end{equation}
where the expressions of the components $q_{i\mathbf{k}}$ and the normalized
factor $G_{n}(\mathbf{k})$ for each band are given in Table I.

\begin{table}[th]
\caption{The expressions for the coefficients in Eq. (\ref{wave}) with $x_{i}%
$=$\mathbf{k\mathtt{\cdot}a}_{i}$.}
%\label{specs}%
\begin{tabular}
[c]{cc}\hline\hline
$q_{1k}$ & $\ \ \ \ \ \ \ \ \ \ \ \ \ \ \ \ \ \ \frac{1}{2}[\epsilon
_{nk}^{(0)2}-4\cos^{2}x_{2}]$\\
$q_{2k}$ & $\ \ \ \ \ \ \ \ \ \ \ \ \ \ \ \ \ \ \epsilon_{nk}^{(0)}\cos
x_{1}+2\cos x_{2}\cos x_{3}$\\
$q_{3k}$ & $\ \ \ \ \ \ \ \ \ \ \ \ \ \ \ \ \ \ \epsilon_{nk}^{(0)}\cos
x_{3}+2\cos x_{2}\cos x_{1}$\\
$G_{nk}^{-2}$ & $\ \ \ \ \ 2b_{k}\epsilon_{nk}^{(0)2}+[4b_{k}-3\epsilon
_{nk}^{(0)2}]\cos^{2}x_{2}+6(b_{k}-1)\epsilon_{nk}^{(0)}\ \ $\\\hline
\end{tabular}
\end{table}

Then, we introduce the intrinsic SOI term, which, according to the symmetry of
the kagom\'{e} lattice, takes the form \cite{Kane1,Guo}
\begin{equation}
\mathcal{H}_{\text{SO}}=i\frac{2\lambda_{\text{SO}}}{\sqrt{3}}\sum
_{\langle\langle ij\rangle\rangle\sigma_{1}\sigma_{2}}\left(  \mathbf{d}%
_{ij}^{1}\times\mathbf{d}_{ij}^{2}\right)  \cdot\mathbf{s}_{\sigma_{1}%
\sigma_{2}}c_{i\sigma_{1}}^{\dag}c_{j\sigma_{2}}. \label{Hso}%
\end{equation}
Here $\lambda_{\text{SO}}$ represents the SOI strength, $\mathbf{s}$ is the
vector of Pauli spin matrices, $i$ and $j$ are next-nearest neighbors, and
$\mathbf{d}_{ij}^{1}$ and $\mathbf{d}_{ij}^{2}$ are the vectors along the two
bonds that connect $i$ to $j$. Taking the Fourier transform, we have
\[
\mathcal{H}_{\text{SO}}\mathtt{=}\sum_{\mathbf{k}\sigma}\psi_{\mathbf{k}%
\sigma}^{+}H_{\text{SO}}(\mathbf{k})\psi_{\mathbf{k}\sigma}%
\]
with%
\begin{equation}
H_{\text{SO}}(\mathbf{k})=\pm2\lambda_{\text{SO}}\left(
\begin{array}
[c]{ccc}%
0 & i\cos(\mathbf{k}\mathtt{\cdot}\mathbf{b}_{1}) & -i\cos(\mathbf{k}%
\mathtt{\cdot}\mathbf{b}_{3})\\
-i\cos(\mathbf{k}\mathtt{\cdot}\mathbf{b}_{1}) & 0 & i\cos(\mathbf{k}%
\mathtt{\cdot}\mathbf{b}_{2})\\
i\cos(\mathbf{k}\mathtt{\cdot}\mathbf{b}_{3}) & -i\cos(\mathbf{k}%
\mathtt{\cdot}\mathbf{b}_{2}) & 0
\end{array}
\right)  , \label{HsoK}%
\end{equation}
where $\mathbf{b}_{1}$=$\mathbf{a}_{3}-\mathbf{a}_{2}$, $\mathbf{b}_{2}%
$=$\mathbf{a}_{1}-\mathbf{a}_{3}$, $\mathbf{b}_{3}$=$\mathbf{a}_{2}%
-\mathbf{a}_{1}$, and the $+$($-$) sign refers to spin up (down) electrons.

\begin{figure}[ptb]
\begin{center}
\includegraphics[width=0.5\linewidth]{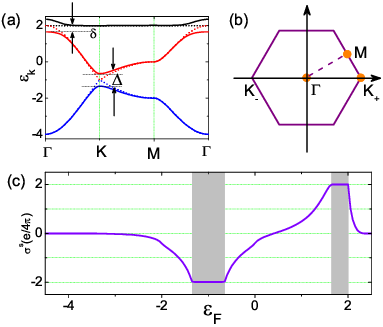}
\end{center}
\caption{(Color online). (a) Energy spectrum (solid curves) of the 2D
kagom\'{e} lattice along the high-symmetry lines in the BZ with intrinsic
spin-orbit couplings $\lambda_{\text{SO}}$=$0.1t$. There are two band gaps
appearing with gap width $\Delta$ and $\delta$. For comparison, we also draw
the energy spectrum without intrinsic spin-orbit couplings (dotted lines). (b)
The corresponding BZ of the 2D kagom\'{e} lattice. (c) The SHC $\sigma
_{xy}^{s}$ as a function of the Fermi energy $\epsilon_{F}$. The shaded areas
correspond to the bulk gaps.}%
\label{f2}%
\end{figure}

Inclusion of the intrinsic SOI in the Hamiltonian makes the
appearance of the eigenstates $\left\vert
u_{n\mathbf{k}}\right\rangle $ and eigenenergies
$\epsilon_{n\mathbf{k}}$ very tedious with exception at some
high-symmetry $\mathbf{k}$ points. Instead of writing their explicit
forms, here we show in Fig. \ref{f2}(a) (solid curves) the
numerically calculated energy spectrum for
the total Hamiltonian $\mathcal{H}\mathtt{=}\mathcal{H}_{0}\mathtt{+}%
\mathcal{H}_{\text{SO}}$ along the high-symmetry lines
($\mathbf{\Gamma }\mathtt{\rightarrow}\mathbf{K}$,
$\mathbf{K}\mathtt{\rightarrow}\mathbf{M}$, and
$\mathbf{M}\mathtt{\rightarrow}\mathbf{\Gamma}$) in the BZ. The SOI\
coefficient is chosen to be $\lambda_{\text{SO}}$=$0.1t$. For
comparison we also plot in Fig. \ref{f2}(a) (dashed curves) the
energy spectrum in the absence of SOI ($\lambda_{\text{SO}}$=$0$).
One can see that while the spin degeneracy is not lifted by the
presence of the intrinsic SOI, nevertheless, the Dirac contacts of
the lower and middle bands at the $\mathbf{K}$ point are removed and
a gap of amplitude $\Delta$=$4\sqrt{3}|\lambda_{\text{SO}}|$ opens
between these two bands. The amplitude of this gap turns out to be
$\Delta $=$4\sqrt{3}|\lambda_{\text{SO}}|$. Similarly, the original
contact at the $\mathbf{\Gamma}$ point between the middle and upper
(flat) bands is also lifted by the presence of SOI and a gap with
amplitude $\delta$ is opened. However, this gap is an indirect one,
i.e., the middle-band maximum and upper-band minimum are not at the
same $k$-point.

To see the behavior of the system in the insulating state, we have calculated
the SHC using the following Kubo formula \cite{Sinova}%
\begin{align}
\sigma_{xy}^{s}  &  =-e\hslash\sum_{n\neq n^{\prime},\mathbf{k}}\left[
f\left(  \epsilon_{n\mathbf{k}}\right)  -f\left(  \epsilon_{n^{\prime
}\mathbf{k}}\right)  \right] \label{kubo}\\
&  \times\frac{\operatorname{Im}\langle u_{n\mathbf{k}}|\frac{1}{2}\left\{
\hat{v}_{x},\hat{s}_{z}\right\}  |u_{n^{\prime}\mathbf{k}}\rangle\langle
u_{n^{\prime}\mathbf{k}}|\hat{v}_{y}|u_{n\mathbf{k}}\rangle}{\left(
\epsilon_{n\mathbf{k}}-\epsilon_{n^{\prime}\mathbf{k}}\right)  ^{2}+\eta^{2}%
},\nonumber
\end{align}
where $\mathbf{\hat{v}}\left(  \mathbf{k}\right)  $=$\partial H(\mathbf{k}%
)/\hslash\partial\mathbf{k}$ and $H(\mathbf{k})$=$H_{0}(\mathbf{k}%
)+H_{\text{SO}}(\mathbf{k})$. The calculated result at zero
temperature is shown in Fig. \ref{f2}(c) by varying the Fermi energy
$\epsilon_{F}$ with the SOI\ coefficient
$\lambda_{\text{SO}}$=$0.1t$. From Fig. 2(c), one can see that
initially the SHC $\sigma_{xy}^{s}$ decreases as the filling factor
of the (spin-degenerate) lower band increases, arriving at the
minimum value $-(e/2\pi)$ at $\epsilon_{F}$=$-1.35$
(=$-1\mathtt{-}\Delta/2$), a value corresponding to the top of the
lower band. Then, as the Fermi energy $\epsilon_{F}$\ continues to
vary in the first gap region (shaded area wherein
$-1.35\mathtt{\leqslant}\epsilon_{F}\mathtt{\leqslant}-0.65$), the
SHC keeps this minimum value unchanged. As shown in the following
discussion, this quantized SHC can be understood by the
\textbf{Z}$_{2}$-valued topological invariant associated with this
quantum spin Hall phase \cite{Kane2}. When the Fermi
energy increases to touch the bottom of the middle band at $\epsilon_{F}%
$=$-0.65$(=$-1+\Delta/2$), then the SHC suddenly switches up and rapidly
increases when the Fermi energy goes through the middle two bands. When
$\epsilon_{F}$ increases to be at the top of the middle two bands ($\epsilon_{F}%
$=$1.65$), $\sigma_{xy}^{s}$ arrives at the maximum value $e/2\pi$. Then as
the Fermi energy $\epsilon_{F}$\ continues to vary in the second gap region
(shaded area $1.65\mathtt{\leqslant}\epsilon_{F}\mathtt{\leqslant}2$), the SHC
$\sigma_{xy}^{s}$ keeps this maximum value unchanged. When the Fermi energy
increases to touch the bottom of the upper band at $\epsilon_{F}$=$2$, the SHC
then suddenly switches down and rapidly decreases when the Fermi energy goes
through the upper band. Finally the SHC $\sigma_{xy}^{s}$ decreases to
disappear when the three spin-degenerate bulk bands are all fully occupied.

Topologically, the quantum spin Hall insulating state at $1/3$ or $2/3$
filling can be seen by calculating a selective \textbf{Z}$_{2}$-valued
invariant $\nu$ \cite{Kane2}, which is related to the parity eigenvalues
$\xi_{2m}(\Gamma_{i})$ of the $2m$-th occupied energy band at the four
time-reversal-invariant momenta $\Gamma_{i}$ \cite{Fu}. Very recently, Guo and
Franz \cite{Guo} have numerically calculated the eigenstate of $\mathcal{H}%
_{\Gamma_{i}}$ and they found that three $\xi$'s are positive and one is
negative. Although which of the four $\xi$'s is negative depends on the choice
of the inversion center, the product $\Pi_{i}\xi(\Gamma_{i})$=$(-1)^{\nu}$ is
independent of this choice and determines the nontrivial \textbf{Z}$_{2}$
invariant $\nu$=$1$. That confirms the kagom\'{e} lattice system to be a
quantum spin Hall TBI at $1/3$ (or $2/3$) filling.

\begin{figure}[ptb]
\begin{center}
\includegraphics[width=0.5\linewidth]{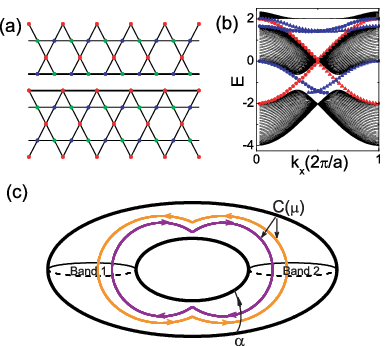}
\end{center}
\caption{(Color online). (a) Sketch of the 2D kagom\'{e} lattice strip with
two edges along the $y$ direction. The red, blue, and green dots are used to
distinguish the independent sites A, B, and C, respectively. (b) Energy
spectrum of the kagom\'{e} lattice strip ($N_{y}$=$40$) with the Hamiltonian
$\mathcal{H}$=$\mathcal{H}_{0}+\mathcal{H}_{\text{SO}}$. The spin-orbit
coupling strength is set as $\lambda_{\text{SO}}$=$0.1t$. The red and blue
lines represent the edge states localized at the down and up edges of the
system, respectively. And the circle and triangle label the up and down spins,
respectively. (c) The Riemann surface of the Bloch function corresponding to
$1/3$ filling. The purple and orange curves correspond to spin-up and
spin-down channels, respectively. }%
\label{f32}%
\end{figure}

On the other hand, the topological aspect of the quantized spin Hall phase can
be distinguished by the difference between the winding numbers of the spin-up
and spin-down edge states across the holes of the complex-energy Riemann
surface, $I_{s}$=$I_{\uparrow}-I_{\downarrow}$ \cite{Wang5}. The SHC is then
given by $\sigma_{xy}^{s}$=$I_{s}\left(  e/4\pi\right)  $. Using this
topological index $I_{s}$ we have investigated the quantum spin Hall effect in
the Kane-Mele graphene model \cite{Wang5}. For the present kagom\'{e} lattice
model we can also study the quantum spin Hall effect in terms of this
topological winding index. For this purpose, let us first numerically
diagonalize the total Hamiltonian $\mathcal{H}$ using the strip geometry. For
convenience and without loss of generality, we suppose the system has two
edges in the $y$ direction while keeping infinite in the $x$ direction [see
Fig. \ref{f32}(a)]. The number of sites $A$ (or $B,C$) in the $y$ direction is
chosen to be $N_{y}$=$40$. The calculated energy spectrum is drawn in Fig.
\ref{f32}(b). From this figure one can clearly see that there are spin edge
states occurring in each energy gap. These gapless edge states in the
truncated kagom\'{e} lattice are topologically stable against random-potential
perturbation, provided that the perturbation is small compared to the bulk
gaps. We have numerically confirmed this fact. The Riemann surface of Bloch
function is plotted in Fig. \ref{f32}(c) for $1/3$ filling. According to Ref.
\cite{Wang5}, the winding number of spin-up (spin-down) edge state in the
lower gap is $I_{\uparrow}$=$-1$ ($I_{\downarrow}$=$1$), which gives $I_{s}%
$=$-2$ at $1/3$ filling. That means the SHC in this phase is quantized as
$\sigma_{xy}^{s}$=$-(e/2\pi)$. The Riemann surface of Bloch function at $2/3$
filling are the same as that at $1/3$ filling, except that the directions of
the curves corresponding to different spin channels are inverse. So the
winding number of spin-up (spin-down) edge state in the upper gap is
$I_{\uparrow}$=$1$ ($I_{\downarrow}$=$-1$), which gives $I_{s}$=$2$. The
corresponding SHC at $2/3$ filling is then quantized as $\sigma_{xy}^{s}%
$=$e/2\pi$. These conclusions are consistent with those calculated by using
the Kubo formula (\ref{kubo}) [see Fig. \ref{f2}(c)]. Note that although the
non-trivial \textbf{Z}$_{2}$ invariant $\nu$=$1$ in the bulk analysis can
confirm the quantum spin Hall TBI phase, it does not provide the information
on the sign of the SHC. In contrast, our winding-number analysis can resolve
this sign at different fillings (i.e., $\sigma_{xy}^{s}$=$\mp(e/2\pi)$ at
$1/3$ and $2/3$ filling, respectively).

The topological properties of the system in the insulating state are stable
even when the Rashba SOI is considered. To clearly see this fact, we
numerically calculated the SHC $\sigma_{xy}^{s}$ with the Kubo formula
(\ref{kubo}) when the total Hamiltonian includes the Rashba SOI \cite{Liu}%
\begin{equation}
\mathcal{H}_{R}=i\frac{\lambda_{R}}{\hslash}\sum_{\langle ij\rangle\sigma
_{1}\sigma_{2}}c_{i\sigma_{1}}^{\dag}(\mathbf{s}\times\mathbf{\hat{d}}%
_{ij})_{z}c_{j\sigma_{2}}, \label{Hrashba}%
\end{equation}
where $\lambda_{R}$ is the Rashba coefficient and $\mathbf{\hat{d}}_{ij}$ is a
vector along the bond the electron traverses going from site $j$ to $i$. The
calculated SHC is drawn in Fig. 4 for different values of $\lambda_{R}$.
Clearly, one can see that the amplitude of SHC in the insulating phase keeps
$e/2\pi$ unchanged when the Rashba SOI strength $0\mathtt{\leqslant}%
\lambda_{R}/t\mathtt{<}0$.$1$. When the Rashba SOI is sufficiently large (for
example $\lambda_{R}$=$0$.$2t$), the amplitude of SHC will depart little from
the quantized value. However, the topology of this insulating state keeps
unchanged, unless the bulk gaps disappear.

\begin{figure}[ptb]
\begin{center}
\includegraphics[width=0.5\linewidth]{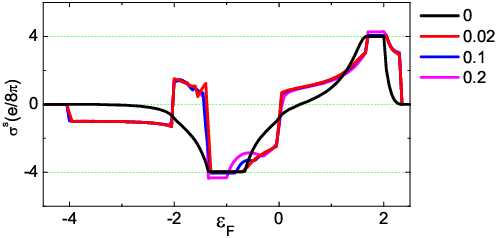}
\end{center}
\caption{(Color online). The SHC as a function of the Fermi energy
$\epsilon_{F}$ with different values of the Rashba coefficient. The black,
red, blue, and pink lines correspond to $\lambda_{R}$=$0$, $0.02t$, $0.1t$,
and $0.2t$, respectively. The intrinsic SOI strength is set as $\lambda
_{\text{SO}}$=$0.1t$.}%
\end{figure}

\begin{figure}[ptb]
\begin{center}
\includegraphics[width=1.0\linewidth]{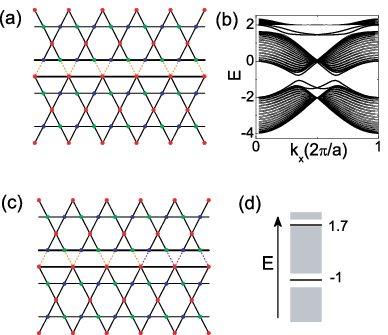}
\end{center}
\caption{(Color online). (a) Two edges of the 2D kagom\'{e} lattice are
reconnected with weaker bonds (orange dotted lines) and its corresponding
energy spectrum (b). (c) The restored bonds have a sign reversal along the
right half bonds (violet dotted lines) and its corresponding energy spectrum
(d). In these figures, the system size is set as $N_{y}$=$40$ and the
spin-orbit coupling $\lambda_{\text{SO}}$=$0.1t$. The bonds connected two
edges are set as $0.25$ times of other ones.}%
\label{f41}%
\end{figure}

In the following let us consider the spin-charge separation in the 2D
kagom\'{e} lattice. If we reconnect the two edges of the kagom\'{e} lattice
shown in Fig. \ref{f32}(a) with weaker bonds, two small gaps then reappear in
the edge spectrum [see Figs. \ref{f41}(a) and (b)]. Topological excitations
(edge solitons) are created by reversing the sign of the reconnected bonds
along the right half row [see Fig. \ref{f41}(c)]. In this manner two defects
with $\pi$ flux are introduced in the present kagom\'{e} lattice. As a result,
four degenerate in-gap spin states localized around these two defects are
formed in each bulk gap [see Figs. \ref{f41}(d)]. The corresponding energies
of these in-gap states are $\epsilon$=$-t$ and $1.7t$, respectively. Note that
in the square lattice version of the Kane-Mele model \cite{Ran}, the in-gap
modes are precisely at zero energy, while in the present kagom\'{e} lattice
model the in-gap modes are no longer at zero energy. Here, we would like to
point out that the zero level discussed in a large amount of previous studies
is a result of particle-hole symmetry. Its role in leading to fermion number
fractionalization was initially found by Jackiw and Rebbi \cite{Jack}, and
then stressed in various insulating systems \cite{Su1980,Lee,Hou,Ran,Qi,Weeks}%
. In this case, a simple physical picture of fractional charge of $e/2$ around
the defect is ready to obtain by the combined fact that (i) under
particle-hole symmetry the relative charge density $\rho$ on the soliton and
the chemical potential $\mu$ satisfy $\rho(\mu)$\texttt{=}$\mathtt{-}\rho
(-\mu)$, and (ii) when $\mu$ is in the bulk gap, the only difference between
$\rho(\mu)$ and $\rho(-\mu)$ is the filling of the zero modes localized at the
two defects. In the present 2D kagom\'{e} lattice, on the other hand, the
system has no particle-hole symmetry and thus no zero mode. In this case,
although we cannot resort to the above simple physical picture, the presence
of soliton-antisoliton doublet (quadruplet when spin included) in each bulk
gap in Fig. 5(d) still guarantees the occurrence of fractionalized excitations.

\begin{figure}[ptb]
\begin{center}
\includegraphics[width=0.8\linewidth]{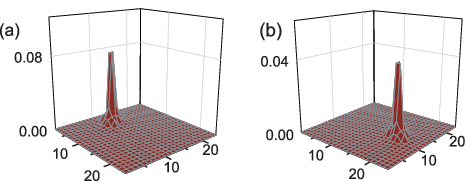}
\end{center}
\caption{(Color online). (a) The charge density of a chargeon state and (b)
the spin density of a spinon on a $24\times24$ lattice with periodic boundary
condition. The chargon at coordinate ($6$, $6$) is $f_{+(1/2)\uparrow
,+(1/2)\downarrow}$, it has charge $-e$ and spin $S_{z}$=$0$. The spinon at
coordinate ($18$, $18$) is $f_{+(1/2)\uparrow,-(1/2)\downarrow}$, it has
charge $0$ and spin $S_{z}$=$\hslash/2$.}%
\label{f51}%
\end{figure}

At 1/3 or 2/3 filling, occupation (unoccupation) of these in-gap modes leads
to an excess (deficit) of $1/2$ fermion number per spin and per defect
\cite{Zhang}. Four different types of solitons with the following quantum
numbers are obtained when these in-gap modes are occupied by different ways:
the chargeon $f_{+(1/2)\uparrow,+(1/2)\downarrow}$ (charge $-e$, $S_{z}$=$0$);
the holon $f_{-(1/2)\uparrow,-(1/2)\downarrow}$ (charge $e$, $S_{z}$=$0$); the
two spinons $f_{+(1/2)\uparrow,-(1/2)\downarrow}$ (charge $0$, $S_{z}$%
=$\frac{\hslash}{2}$) and $f_{-(1/2)\uparrow,+(1/2)\downarrow}$ (charge $0$,
$S_{z}$=$-\frac{\hslash}{2}$). Here the subscript $+$ ($-$) represents that
the in-gap mode is filled (empty) and $\uparrow$ ($\downarrow$) labels the
spin mode. For example, $f_{+(1/2)\uparrow,-(1/2)\downarrow}$ means that the
up-spin mode is filled and the down-spin mode is empty. For further
illustration, the charge-density distribution for the chargeon
$f_{+(1/2)\uparrow,+(1/2)\downarrow}$ is calculated and shown in Fig.
\ref{f51}(a), while the spin-density distribution for the spinon
$f_{+(1/2)\uparrow,-(1/2)\downarrow}$ is plotted in \ref{f51}(b). Here, we use
a $24\mathtt{\times}24$ lattice for calculation.

The quantum statistics of these spin-charge separated solitons can
be readily seen by using anyon fusion argument \cite{Weeks,Ran},
which is based on the observation that the bound states of
fractional excitation acquire non-trivial Berry phases on adiabatic
exchange. Since the spin-up and spin-down bands in the present case
decouple, so without loss of generality, let us consider a bound
state of two identical spin-up solitons $f_{(1/2)\uparrow}$ (or
$f_{-(1/2)\uparrow}$), which carries charge $-e$ ($e$) and flux
$2\pi \mathtt{\sim}0$ thus is a fermion. According to the anyon
fusion rule, which states that the exchange phase $\Theta$\ of a
particle formed by combining $n$ identical anyons with exchange
phase $\theta$ is $\Theta$=$n^{2}\theta$, one easily obtains that
the exchange phase between two identical solitons
$f_{(1/2)\uparrow}$ ($f_{-(1/2)\uparrow}$) should be $1/4$ that of
fermions, i.e., $\theta(f_{(1/2)\uparrow},f_{(1/2)\uparrow})$ (or
$\theta (f_{-(1/2)\uparrow},f_{-(1/2)\uparrow}$)=$\pm\pi/4$. Next
let us consider a bound state of an $f_{(1/2)\uparrow}$ and
$f_{-(1/2)\uparrow}$ soliton, which carries charge $0$ and should be
a boson. Then the exchange phase between solitons
$f_{(1/2)\uparrow}$ and $f_{-(1/2)\uparrow}$ is given by
$\theta(f_{(1/2)\uparrow},f_{-(1/2)\uparrow})$=$\mp\pi/4$. Since the
spin-down band is the Hermitian conjugate of the spin-up band, one
obtains $\theta
(f_{\alpha_{1}\downarrow},f_{\alpha_{2}\downarrow})$=$-\theta(f_{\alpha
_{1}\uparrow},f_{\alpha_{2}\uparrow})$. Substituting these results
into the formula calculating the exchange phase between spin-charge
separated solitons
$\theta(f_{\alpha_{1}\uparrow\beta_{1}\downarrow},f_{\alpha_{2}\uparrow
\beta_{2}\downarrow})$=$\theta(f_{\alpha_{1}\uparrow},f_{\alpha_{2}\uparrow}%
)$+$\theta(f_{\beta_{1}\downarrow},f_{\beta_{2}\downarrow})$, one immediately
concludes that the spinons, holon, and chargeon are all bosons. However, each
spinon has nontrivial mutual exchange phase $\pi$ with the chargeon and holon.

\begin{figure}[ptb]
\begin{center}
\includegraphics[width=1.0\linewidth]{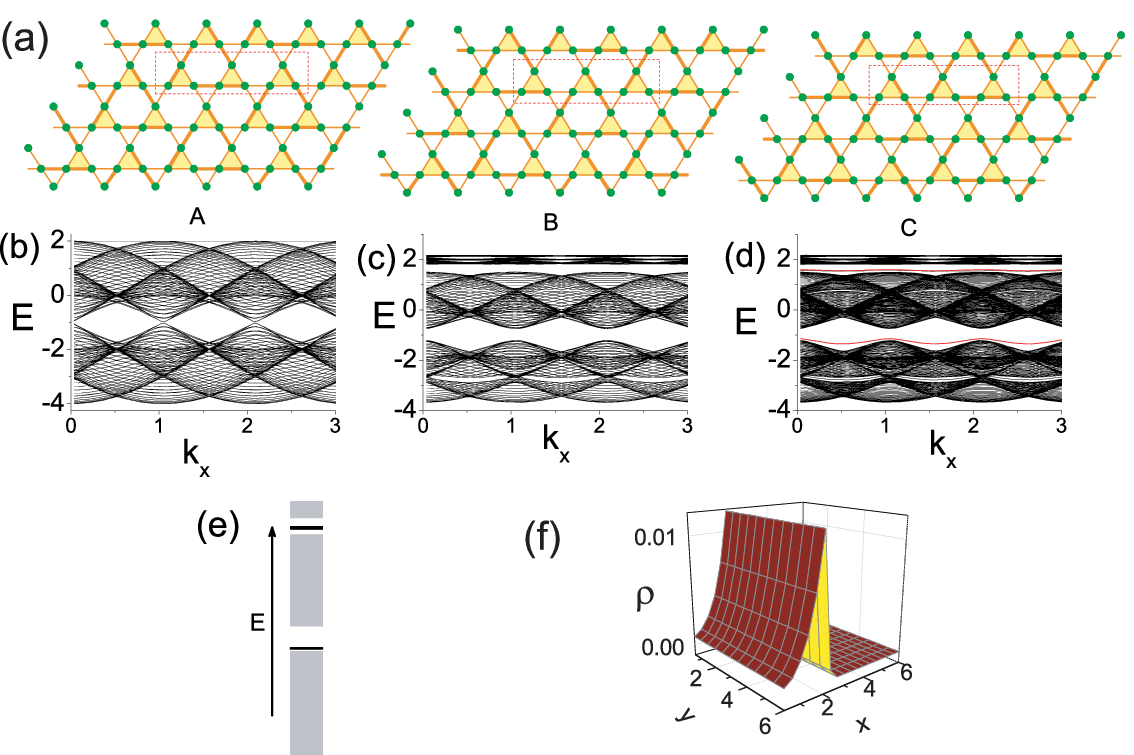}
\end{center}
\caption{(Color online). (a) Schematic pictures of the three
degenerated ground states of a trimerized 2D distorted kagom\'{e}
lattice. The hopping amplitude along the thick (thin) lines is set
as $t+\eta$ ($t-\eta$), where $t$=$-1$ is set as energy unit and
$\eta$=$0.25t$. The dashed lines represent the unit cell, which
contains $9$ lattice sites. The corresponding energy spectrums are
shown in (c) (without boundary) and (d) (with boundary),
respectively. The thick lines in (d) represent the edge states. The
energy spectrum of the undistorted 2D kagom\'{e} lattice
($\eta$=$0$) is drawn in (b). (e) Energy spectrum of the 2D
kagom\'{e} lattice with kink I. The gray shaded regions represent
the bulk bands and the thick black lines represent the excitation
energies. (f) The charge density of a fractional chargeon state on a
trimerized kagome lattice composed by $36$ units. The chargeon at
$x$=$2$ line has charge
$-2e/3$ and spin $S_{z}$=$0$.}%
\label{Fig7}%
\end{figure}

Before ending this paper, we would like to point out that there are
other ways for creating the spin-charge separated solitons in the
kagom\'{e} lattice. For example, by trimerizing the kagom\'{e}
lattice \cite{Guo}, like the 1D way proposed by Su and Schrieffer
\cite{Su}, the excitations of possessing fractional charge $\pm e/3$
or $\pm2e/3$ in two spacial dimensions can be realized. To make this
point more clear, now we construct a trimerized pattern of the
kagom\'{e} lattice by introducing an appropriate bond distortion as
shown in Fig. \ref{Fig7}(a). The tight-binding Hamiltonian
describing this
trimerized kagom\'{e} lattice is generally written as $\mathcal{H}_{0}%
=\sum_{\langle ij\rangle}t_{ij}c_{i}^{\dag}c_{j}$, where the hopping
amplitude $t_{ij}$ are different for different nearest-neighbor link
$\langle ij\rangle$ when the distortion is introduced. For
simplicity and without loss of generality, we set in Fig.
\ref{Fig7}(a) the hopping term along the thick (thin) bonds as
$t+\eta$ ($t-\eta$), where $|\eta|$ depicts the distortion
amplitude, which is smaller than the undistorted hopping amplitude
$|t|$. In this case, the unit cell now becomes larger and contains
$9$ sites [see the dashed lines in Fig. \ref{Fig7}(a)]. Clearly,
there are three different phases corresponding to the ground state
(denoted by A, B, and C, respectively). The corresponding spectrum
of this distorted kagom\'{e} lattice (A, B, or C) with infinite size
is drawn in Fig. \ref{Fig7}(c). When the system is finite, i.e., it
has two boundaries along one direction (say, the $x$ direction),
there are edge states appearing in the bulk gaps [see the red thick
lines in the spectrum drawn in Fig. \ref{Fig7}(d)]. For comparison,
we also plot in Fig. \ref{Fig7}(b) the spectrum of the 2D perfect
undistorted kagom\'{e} lattice. In this case the 2D kagom\'{e}
lattice system has no bulk gaps. Two facts should be noted. One is
that the bulk gaps appearing in Figs. \ref{Fig7}(c) and 7(d) now
result from the distortion. The other is that the in-gap edge states
in Fig. \ref{Fig7}(d) do not connect the neighboring two bulk bands.
According to the winding properties of the edge states
\cite{Hatsugai,Wang5}, one knows that the insulating phase risen
from the distortion is topologically trivial. That means the Hall
conductance is zero when the Fermi energy lies in the bulk gaps.
This point has been validated by the numerical calculation with the
linear-response formula \cite{Sinova}.

We can now study the structure of kinks connecting different phases.
Similar to the case of one spatial dimension \cite{Su}, we
distinguish two classes of kinks: type I, which leads from A to B, B
to C, or C to A as one moves from left to right; and type II, which
leads from A to C, C to B, or B to A as one moves along the same
direction. Fig. \ref{Fig7}(e) plots the energy spectrum of the
trimerized kagom\'{e} lattice with kink I (or with kink II) by using
$1296$ sites. Every phase has $6\times8$ units and in each unit
there are $9$ independent sites. For numerical calculation, we set
$t$=$-1$ and the distortion $\eta $=$0.25t$. From Fig. \ref{Fig7}(e)
one can clearly observe that there are excitation energies appearing
in the bulk gaps. In this case, in the lower gap there are $9$
excitation energies while in the higher gap there are $48$ ones.
These excitations are localized around the kinks, which is shown in
Fig. \ref{Fig7}(f). Every kink has fractional charge $-e/3$. That
means we have realized the excitations possessing fractional charge
$-e/3$ in the trimerized kagom\'{e} lattice. When we numerically
increase the system size along the $y$ axis, the number of the
excitation energies is found to increase in proportion to the system
size. In the infinite limit, an excitation band eventually forms in
the bulk gap. However, when we increase the size along the $x$ axis,
the number of the excitations keeps unchanged. When the spin freedom
is considered, one can obtain the spin-charge separated excitations
at $1/3$ (or $2/3$) filling. For example, in a small system with
$6\times6$ units, there are $4$ double-degenerate excitation states
lying in the lower gap, one half is contributed from the lower bands
and one half from the middle one. According to the different
occupation ways, the spin-charge separated excitations with
fractional quantum numbers, chargeon $\pm2e/3$, as well as other
excitations with fractional charge $\pm e/3$ and spin
$\pm\hslash/2$, are obtained.

In summary, we have theoretically studied quantum spin Hall effect
and spin-charge separation in a 2D kagom\'{e} lattice. By using the
topological winding numbers of the spin edge states on the
complex-energy Riemann surface, we have obtained that the SHC is
quantized as $\pm(e/2\pi)$ when the system is in the insulating
phases, which is consistent with the \textbf{Z}$_{2}$ topological
invariant analysis and the numerical linear-response calculation.
Furthermore, we have constructed the spin-charge separated solitons
in the kagom\'{e} lattice by connecting the system's boundaries in
twist. The quantum statistics of these solitons has also been
discussed.

This work was supported by NSFC under Grants No. 10604010, No. 10904005, and
No. 60776063, and by the National Basic Research Program of China (973
Program) under Grant No. 2009CB929103.

\end{document}